# Strong frequency correlation and anti-correlation between a Raman laser and its pump laser for positive and negative dispersions

Zifan Zhou[1,*], Ruoxi Zhu[1], Selim M. Shahriar[1,2]

[1]Department of Electrical and Computer Engineering, Northwestern University, Evanston, IL 60208, USA

[1]Department of Physics and Astronomy, Northwestern University, Evanston, IL 60208, USA

[*]zifan.zhou@northwestern.edu

**Abstract:** We show that the frequency of a Raman laser is highly correlated or anti-correlated with the frequency of the Raman pump laser, depending on whether the dispersion experienced by the Raman laser is positive or negative. For a subluminal laser, corresponding to a positive dispersion with a group index that is much larger than unity, the shift in its frequency is approximately the same as that in the Raman pump laser. In contrast, for a superluminal laser, corresponding to a negative dispersion with a group index that is close to zero, its frequency shifts in the direction opposite to that of the Raman pump lasers, and has an amplitude that is larger by a factor approximately equaling the inverse of the group index. These findings would play a critical role in determining the maximum achievable sensitivity of sensors employing such lasers, especially under conditions where the pump laser linewidth is broadened significantly beyond the Schawlow-Townes linewidth due to classical fluctuations.

## 1. Introduction

For a highly dispersive laser (HDL), the change in the frequency as a function of a change in the cavity length or other mechanisms, such as the Sagnac effect, gets greatly amplified or suppressed [1,2,3], compared to a conventional laser, depending on the group index. This property makes HDLs potentially suitable for precision metrology, such as rotation sensing [4,5,6,7,8], gravitational wave detection [9,10], and dark matter search [11,12]. HDLs can be divided into two categories based on their properties: superluminal lasers and subluminal lasers. The superluminal (subluminal) laser is a laser inside which the group velocity of the laser field is much larger (smaller) than the speed of light in vacuum. In order to realize an HDL, it is necessary for the gain spectrum to be much narrower than what is typically used for lasers. In principle, many different approaches can be used to realize such gain profiles. However, for many applications, including rotation sensing, it is necessary to ensure that the gain process is unidirectional. As such, in our recent studies of both subluminal [13,14] and superluminal lasers [15,16,17,18,19,20,21,22,23], both theoretically and experimentally, we have made use of Raman transitions in alkali atoms to produce the requisite narrow features in the gain spectrum. For the case of positive dispersion, necessary for subluminal lasers, the gain is produced simply by employing a three-level Λ transition, along with an auxiliary level that enables optical pumping of atoms into one of the two ground states. Application of a pump laser coupling this state to the intermediate state, under large detuning, produces a narrow gain spectrum for a probe applied on the other leg of the Λ transition, around the frequency that corresponds to two-photon resonance [24,25]. For the case of negative dispersion, necessary for superluminal lasers, we have usually combined a pair of Λ transitions in two different isotopes: one for producing a narrow gain peak, and the other for producing a narrower dip in the gain profile. We have investigated many variations of this dual-isotopes approach [17,18,20,21]. We have also investigated other approaches for realizing a superluminal laser employing a single isotope only [16,19,22,23]. In this paper, we only consider the case where two isotopes are used. Specifically, we use the approach presented in Ref. 20 and Ref. 20 for concreteness of discussions. Due to the similarity in the gain and dispersion spectra of the approaches using the dual-isotopes, the behavior of superluminal lasers employing other dual-isotopes based schemes can be somewhat different from the results presented in the paper. On the other hand, the conclusion reached here may not necessarily apply to the schemes employing a single isotope. The exact behavior of each individual approach can be investigated by solving the explicit model that includes all the relevant energy levels and optical transitions.

In most of these studies, we had assumed that the frequency of the pump laser is a delta function. However, in one recent study, we had investigated the effect of fluctuations in the frequency of the pump laser [26]. In this study, we observed experimental evidence of a correlation between the frequency of the Raman laser and that of the pump laser. In order to account for this effect, we investigated a mechanism in which the spectral width of the pump laser is modeled in terms of random jumps in its phase. First, the gain for a weak probe was calculated using the steady-state solutions of the density matrix equations of motion, for a given phase of the pump laser and a given phase of the probe. This gain was independent of the relative phase between the pump and the probe. Next, we changed the phase of the pump abruptly, and determined the temporal variation of the gain over a time window that is short compared to the time it takes for the system to achieve a steady state. The gain was then averaged over this time window, and the averaged gain amplitude was studied as a function of changes in the probe phase. It was found that the averaged gain was maximum when the change in the probe phase matched the change in the pump phase. While this model provided a plausible reason for the observed correlation between the frequencies of the pump and the Raman lasers, it was not rigorous nor complete enough to predict quantitatively the degree of this correlation. In this paper, we describe a model that evaluates the frequency shift in the Raman laser when the center frequency of the Raman pump is varied, while ignoring its spectral width. The model we presented in Ref. 26 was stochastic, and can possibly be used to determine the correlation between the random phase jumps of the pump and the laser. In contrast, the model we present here is deterministic, and can be used to determine the correlation between the shifts in the center frequency of the laser and that of the pump. This analysis would apply to situations where the change in the pump frequency is not due to random phase jumps.

It needs to be noted that for the approach considered in this paper, the cause of the frequency shift in the Raman laser is a frequency shift in the Raman pump, which is formally equivalent to a change in the hyperfine splitting between the two ground states. The change in the hyperfine splitting between the ground states can be caused by many effects, including the light-shifts produced by the Raman pump and the optical pump. The conclusion reached here can be applied to determine the Raman laser frequency shifts induced by a change in the hyperfine energy difference in addition to a frequency shift in the Raman pump.

The rest of the paper is organized as follows. In Section 2, we describe the effect of the perturbation in the Raman pump laser frequency on a subluminal laser with linear index variations. In Section 3, we describe how the superluminal laser behaves as a function of the Raman pump frequency for both linear and non-linear index variations. In Section 4, we discuss the effects of light shifts produced by the Raman pump laser(s), the optical pumping laser(s), as well as the highly dispersive laser (HDL) itself. In Section 5, we discuss the implications of the findings of this paper on HDL based sensors. Concluding remarks are presented in Section 6.

## 2. Subluminal Laser

The simplest subluminal laser is just a Raman laser, employing a three-level atomic system, as illustrated schematically in the left panel of Figure 1. Here, we assume that an optical pump transfers atoms from state $|1\rangle$ to state $|2\rangle$, via coupling to an auxiliary level not shown in this diagram. This process can be represented simply as a decay rate from state $|1\rangle$ to state $|2\rangle$. When this rate is stronger that the rate of collisional excitation from state $|2\rangle$ to state $|1\rangle$, the population in state $|2\rangle$ will be larger than that in state $|1\rangle$ in steady state, in the absence of any other fields. This creates the so-called Raman population inversion, which is needed for Raman gain. For simplicity of discussion and analytical modeling, we ignore here the rate of collisional excitation from state $|1\rangle$ to state $|2\rangle$ and vice versa, and consider only the effective decay rate from state $|1\rangle$ to state $|2\rangle$ due to the optical pumping process.

Consider next the application of a Raman pump along the $|2\rangle \leftrightarrow |3\rangle$ transition and a Raman probe (with vanishingly small intensity) along the $|1\rangle \leftrightarrow |3\rangle$ transition. Here, $\Omega_P$ and $\Omega_L$ are the

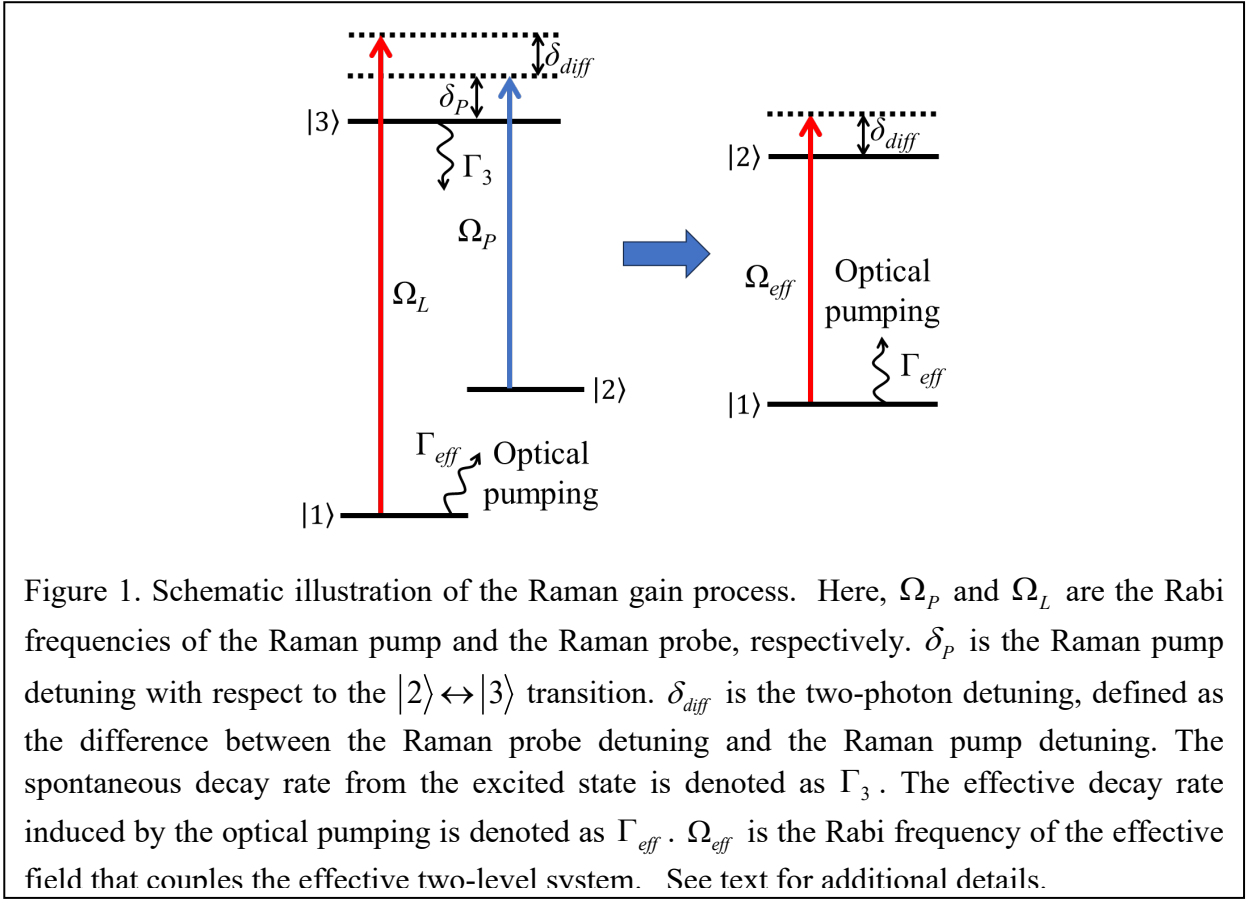

Figure 1. Schematic illustration of the Raman gain process. Here, $\Omega_P$ and $\Omega_L$ are the Rabi frequencies of the Raman pump and the Raman probe, respectively. $\delta_P$ is the Raman pump detuning with respect to the $|2\rangle \leftrightarrow |3\rangle$ transition. $\delta_{diff}$ is the two-photon detuning, defined as the difference between the Raman probe detuning and the Raman pump detuning. The spontaneous decay rate from the excited state is denoted as $\Gamma_3$. The effective decay rate induced by the optical pumping is denoted as $\Gamma_{eff}$. $\Omega_{eff}$ is the Rabi frequency of the effective field that couples the effective two-level system. See text for additional details.

Rabi frequencies of the Raman pump and the Raman probe, respectively. We denote by $\delta_P$ the Raman pump detuning with respect to the $|2\rangle \leftrightarrow |3\rangle$ transition, and by $(\delta_P + \delta_{diff})$ the Raman probe detuning with respect to the $|1\rangle \leftrightarrow |3\rangle$ transition. As such, the quantity $\delta_{diff}$ represents the two-photon detuning, which can also be called the Raman detuning. The Raman gain is maximal when $\delta_{diff} = 0$; thus, the value of $\delta_{diff}$ remains small during the operation of the Raman laser. In contrast, we choose the value of $\delta_P$ to be very large compared to the excited state linewidth (for a Doppler broadened medium, the value of $\delta_P$ is chosen to be larger than the Doppler width of the $|2\rangle \leftrightarrow |3\rangle$ transition). We also assume that $\delta_P \gg \Omega_P \gg \Omega_L$. Under these conditions, one can adiabatically eliminate state $|3\rangle$ [27], and the interaction can be modeled as being equivalent to an effective two level system, as shown in the right panel of Figure 1. The derivations for adiabatic elimination of the excited state are included in **Appendix A**. In this two-level system, the Rabi frequency of the effective field can be expressed as:

$$\Omega_{eff} \approx \frac{\Omega_P \Omega_L}{2\delta_P}, \tag{1}$$

and the detuning is given by the two-photon detuning, $\delta_{diff}$. In steady state, the coherence of this two-level system can be written as:

$$\tilde{\rho}_{21} = \frac{\Omega_{eff}(2\delta - i\Gamma_{eff})}{2\Omega_{eff}^2 + \Gamma_{eff}^2 + 4\delta^2}, \tag{2}$$

where $\Gamma_{eff}$ is the effective decay rate due to optical pumping and $\delta \equiv \delta_{diff} + (\Omega_L^2 - \Omega_P^2)/4\delta_P$. It needs to be noted that the Rabi frequency and the detuning of the Raman pump are kept fixed. Therefore, $\Omega_{eff}$ is proportional to $\Omega_L$, and we can write $\Omega_{eff} = \theta \Omega_L$ where we have defined $\theta \equiv \Omega_P / 2\delta_P$ as a constant parameter. We further define $\Gamma \equiv \Gamma_{eff}$ so that Eq. (2) can be expressed as:

$$\rho_{21} = \frac{\theta \Omega_L (2\delta - i\Gamma)}{2\theta^2 \Omega_L^2 + \Gamma^2 + 4\delta^2}. \tag{3}$$

The polarizability of the gain medium can be then written as:

$$\frac{P}{2} = N\mu_0 \rho_{31}, \tag{4}$$

where $N$ is the number density of atoms and $\mu_0$ is the dipole moment of a single atom for the $|1\rangle \leftrightarrow |3\rangle$ transition. The value of $\rho_{31}$ is approximately proportional to $\rho_{21}$ [28], for the range of parameters of interest, and the ratio $\rho_{31}/\rho_{21}$ is approximately a real number and equivalent to $\theta$, so that we can write $\rho_{31} \approx \theta \rho_{21}$. The validation of this approximation is also presented in **Appendix A**.

We can also express the polarizability in terms of the electric field and the susceptibility as:

$$P = \varepsilon_0 \chi E, \qquad (5)$$

where $\varepsilon_0$ is the vacuum permittivity, $\chi$ is the susceptibility experienced by the subluminal laser field, and $E$ is the amplitude of electric field for the subluminal laser. We can then equate Eq. (4) and Eq. (5) and find:

$$\chi = \frac{2N\theta\mu_0}{\varepsilon_0 E} \rho_{21}. \qquad (6)$$

By definition, we have the Rabi frequency of the subluminal laser as:

$$\Omega_L = \frac{\mu_0 E}{\hbar}. \qquad (7)$$

Equation (6) can be then re-written as:

$$\begin{aligned}
\chi &= \frac{2N\theta\mu_0^2}{\hbar\varepsilon_0 \Omega_L} \rho_{21} \\
&= \frac{2N\mu_0^2}{\hbar\varepsilon_0 \Omega_L} \cdot \frac{\theta\Omega_L(2\delta - i\Gamma)}{2\theta^2\Omega_L^2 + \Gamma^2 + 4\delta^2} \\
&= \frac{2N\theta^2\mu_0^2}{\hbar\varepsilon_0} \cdot \frac{2\delta - i\Gamma}{2\theta^2\Omega_L^2 + \Gamma^2 + 4\delta^2}.
\end{aligned} \qquad (8)$$

The complex index of the gain medium is:

$$\begin{aligned}
n &= \sqrt{1+\chi} \\
&\approx 1 + \chi/2.
\end{aligned} \qquad (9)$$

This approximation is valid for $|\chi| \ll 1$ which is true for utilizing Rb vapor as the gain medium. The unsaturated gain and the index can be then expressed as:

$$G_u = -\frac{\chi''}{2} = \frac{G_0 \Gamma^2/2}{2\theta^2\Omega_L^2 + \Gamma^2 + 4\delta^2}, \qquad (10)$$

$$n_u = 1 + \frac{\chi'}{2} = 1 + \frac{G_0 \Gamma \delta}{2\theta^2\Omega_L^2 + \Gamma^2 + 4\delta^2}, \qquad (11)$$

$$G_0 \equiv \frac{2N\theta^2\mu_0^2}{\hbar\varepsilon_0 \Gamma}, \qquad (12)$$

where the subscription $u$ indicates the unsaturated quantity, and $\chi'$ ($\chi''$) is the real (imaginary) part of the susceptibility. Combining Eq. (7) and Eq. (12), we can express the Rabi frequency as:

$$\Omega_L^2 = \frac{\varepsilon_0 G_0 \Gamma}{2\hbar N \theta^2} E^2. \quad (13)$$

The unsaturated gain and index can be expressed as functions of $E^2$:

$$G_u = \frac{\zeta}{E^2 + \eta}, \quad (14)$$

$$n_u = 1 + \frac{2\zeta\delta/\Gamma}{E^2 + \eta}, \quad (15)$$

$$\zeta \equiv \frac{\hbar N \Gamma}{2\varepsilon_0}, \quad (16)$$

$$\eta \equiv \left(\Gamma^2 + 4\delta^2\right) \frac{\hbar N}{\varepsilon_0 G_0 \Gamma}. \quad (17)$$

It needs to be noted that $\eta$ is a function of the frequency of the subluminal laser, while $\zeta$ is a constant.

The mechanism for a subluminal laser is illustrated schematically in Figure 2(a). For the discussion in this paragraph, the only parts of Figure 2(a) that are relevant are the beams at frequencies $\omega_{P0}$ and $\omega_{L0}$; the remaining parts of this figure are discussed later on. As described earlier, the optical pumping produces a population inversion between the two ground states. When the Raman pump is applied on the $|2\rangle \leftrightarrow |3\rangle$ transition, Raman gain is produced in the vicinity of the two-photon resonance frequency on the $|1\rangle \leftrightarrow |3\rangle$ transition. In the presence of a properly tuned cavity around this medium, this gain produces the subluminal laser. Here, we denote the original frequency of the Raman pump and the subluminal laser as $\omega_{P0}$ and $\omega_{L0}$, respectively. Before the subluminal laser field is established in the cavity, the unsaturated gain and index spectra experienced by a vanishingly weak probe field are illustrated by the solid curves in Figure 2(b), where the horizontal axis is the probe frequency with $\omega_{probe} = \omega_{L0} + \delta$. As can be seen, these spectra shift in frequency when the Raman pump frequency changes. In what follows, we assume that the cavity length is $L_0$, which corresponds to a value for which the original frequency of the subluminal laser coincides with the center frequency of the gain profile.

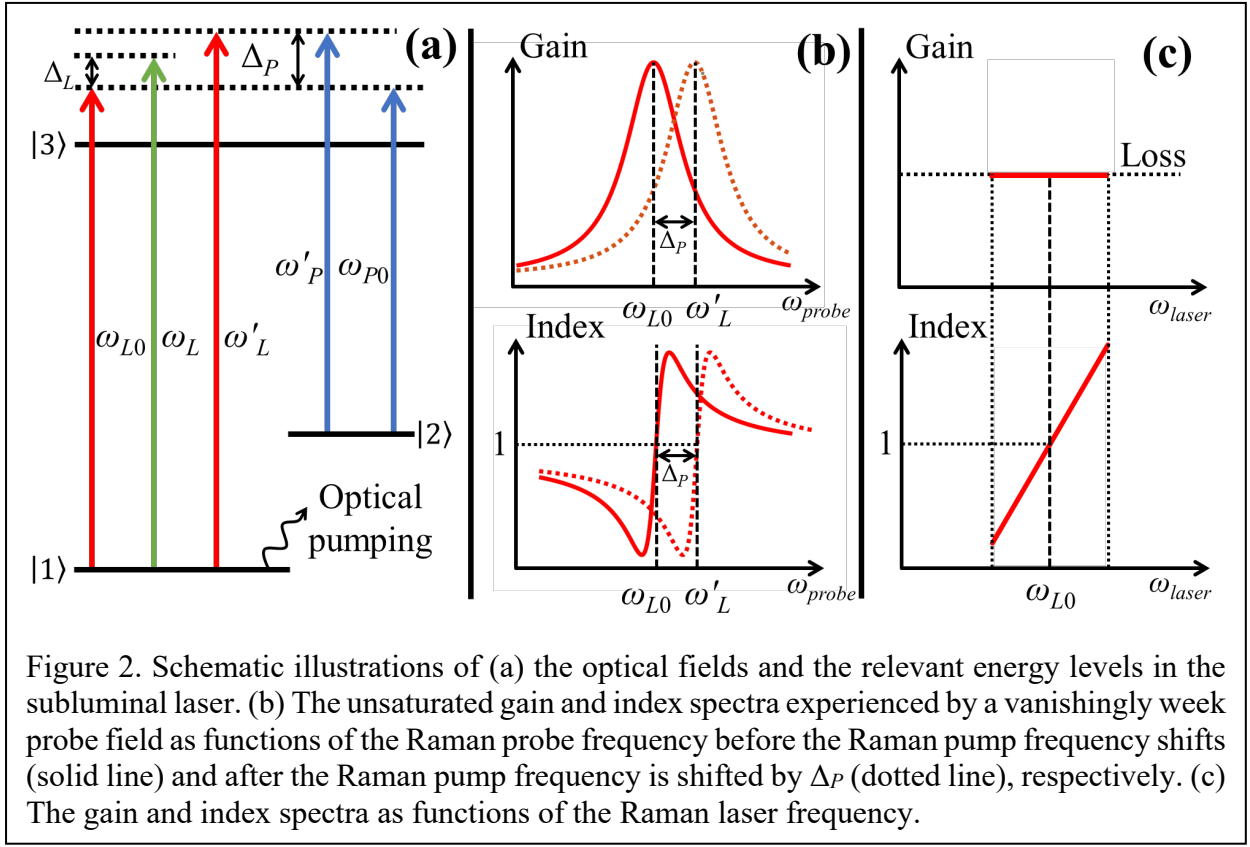

Figure 2. Schematic illustrations of (a) the optical fields and the relevant energy levels in the subluminal laser. (b) The unsaturated gain and index spectra experienced by a vanishingly week probe field as functions of the Raman probe frequency before the Raman pump frequency shifts (solid line) and after the Raman pump frequency is shifted by $\Delta_P$ (dotted line), respectively. (c) The gain and index spectra as functions of the Raman laser frequency.

Next, we consider the situation where the subluminal laser based on this gain is operating, and reaches steady state. The gain and index spectra under this condition are shown in Figure 2(c), where the horizontal axis is the frequency of the subluminal laser, defined as $\omega_{Laser} \equiv \omega_{L0} + \delta$. In the following discussion, we use $\omega$ to replace $\omega_{Laser}$ for simplicity in notations. For concreteness of discussion, we assume that the lasing cavity is a ring resonator consisting of three mirrors, two of which are perfect reflectors, and the output coupler has a finite transmissivity. We also assume that the gain medium fills the whole cavity. For a given cavity length, the lasing frequency will have a fixed value. A variation in the lasing frequency can be produced by changing the cavity length. In these plots, the range of the lasing frequencies is restricted to the values over which the gain experienced by the probe per pass is more than the loss due to the finite reflectivity of the output coupler. As such, the value of the gain shown in the top panel of Figure 2(c), simply equals the loss at the output coupler.

When the cavity length is changed a little, the wavelength of the light has to change in order to ensure that the round-trip phase shift is a multiple of $2\pi$. The corresponding change in the frequency is determined by the index of the gain medium. The profile shown in the bottom panel of Figure 2(c) is based on this interpretation of the index for a laser. By equating the gain to the loss of the cavity, the amplitude of the subluminal laser as a function of the frequency can be determined. Specifically, we equate Eq. (14) to the loss of the cavity, namely $1/2Q$, with $Q$ being the quality factor of the cavity.

$$\frac{\zeta}{E^2+\eta}=\frac{1}{2Q}, \tag{18}$$

$$E^2 = 2Q\zeta - \eta. \tag{19}$$

Using this value of $E^2$ in Eq. (15), we see that the saturated index can be written as:

$$n_s = 1 + \frac{1}{Q\Gamma}\delta, \tag{20}$$

where the subscription $s$ indicates the saturated quantity. As can be seen, the index is perfectly linear over the range of lasing frequencies [15,29], under the assumption that the gain profile for the probe is perfectly Lorentzian. This can be understood as follows. Since the steady-state gain is clamped to a fixed value determined by the transmissivity of the output couple, the intensity of the laser varies with the detuning in a manner that forces the value of $G_s$ to be a constant over this range. Thus, the value of $(n_s - 1)$ becomes simply proportional to the detuning, $\delta$. For simplicity in notations, the index within this frequency range can be expressed as:

$$n_s = 1 + \alpha\delta, \tag{21}$$

where $\alpha \equiv 1/(2Q\Gamma)$ is the slope of the saturated refractive index.

The original frequency of the subluminal laser, $\omega_{L0}$, obeys the phase matching constraint, which can be expressed as:

$$\omega_{L0} n_s = 2\pi m c_0 / L_0. \tag{22}$$

where $m$ is the cavity mode number, $c_0$ is the speed of light in vacuum, and $L_0$ is the fixed length of the laser cavity. As described previously, the cavity length is tuned to the condition where the cavity resonance frequency with the mode number $m$ is the same as $\omega_{L0}$, which yields $n_s(\omega_{L0}) = 1$. The group index at this operating frequency of the subluminal laser is given by:

$$n_g(\omega_{L0}) \equiv \left.\frac{\partial(n_s\omega)}{\partial\omega}\right|_{\omega=\omega_{L0}} = 1 + \alpha\omega_{L0}. \tag{23}$$

Consider next the situation when the frequency of the Raman pump is shifted by an amount of $\Delta_P$, as illustrated by the beam at frequency $\omega'_L$ in Figure 2(a). The center frequency of the unsaturated Raman gain will be shifted by the same amount, as illustrated by the dotted line in the top part of Figure 2(b). The center frequency of the unsaturated index will also be shifted by the same amount, as illustrated by the dotted line in the bottom part of Figure 2(b). We denote the new operating frequency of the subluminal laser as $\omega_L$, as illustrated using the green arrow in Figure 2(a). Our goal is now to determine what the value of $\omega_L$ would be. To determine this, we note first that the subluminal laser frequency that corresponds to unity refractive index, denoted as $\omega'_L$ (which is not necessarily the same as $\omega_L$) is also shifted by $\Delta_P$. The shifted value of the saturated refractive index can be expressed as:

$$n'_s = 1 + \alpha(\omega - \omega'_L). \tag{24}$$

Here we assume that the frequency shift of the Raman pump is small enough so that the slope of the refractive index remains the same, which is equivalent to the width of the gain spectrum remaining unchanged. Since the cavity length remains unchanged, it then follows from the phase-matching condition of Eq. (22) that the new operating frequency of the subluminal laser would obey the relation:

$$\omega_L \cdot n'_s\big|_{\omega=\omega_L} = \omega_{L0} \cdot n_s\big|_{\omega=\omega_{L0}}. \tag{25}$$

where $\omega_L$ is the new frequency of the subluminal laser.

We define the frequency shift in the subluminal laser as:

$$\Delta_L \equiv \omega_L - \omega_{L0}. \tag{26}$$

Using Eq. (25), we have:

$$\omega_L - \omega_{L0} = \omega_{L0}\left(\frac{1}{n'_s\big|_{\omega=\omega_L}} - 1\right). \tag{27}$$

In the expression for $n'_s$ evaluated at $\omega = \omega_L$, we make the assumption that $\alpha(\omega_L - \omega'_L) \ll 1$ which corresponds to the constraint that $(\omega_L - \omega'_L) \ll \Gamma/2G_0$. We can then rewrite Eq. (27) as:

$$\begin{aligned}\omega_L - \omega_{L0} &\approx \omega_{L0}\left\{[1 - \alpha(\omega_L - \omega'_L)] - 1\right\}\\ &= \alpha\omega_{L0}(\omega'_L - \omega_L).\end{aligned} \tag{28}$$

Due to the fact that $\omega'_L - \omega_{L0} = \Delta_P$, we can rearrange Eq. (28) as:

$$\frac{1}{\alpha\omega_{L0}}(\omega_L - \omega_{L0}) = (\omega'_L - \omega_{L0}) + (\omega_{L0} - \omega_L). \tag{29}$$

This can be re-expressed as:

$$\frac{1}{\alpha\omega_{L0}}\Delta_L = \Delta_P - \Delta_L. \tag{30}$$

$$\frac{1+\alpha\omega_{L0}}{\alpha\omega_{L0}}\Delta_L = \Delta_P. \tag{31}$$

From Eq. (23), it then follows that:

$$\frac{\Delta_L}{\Delta_P} \equiv \frac{n_g - 1}{n_g}. \tag{32}$$

For a subluminal laser with a group index much larger than unity, i.e., $n_g \gg 1$, this ratio approaches unity, so that the subluminal laser frequency moves essentially by the same amount as the shift in the Raman pump frequency. When the group index is close to unity, the shift in the subluminal laser frequency can be vanishingly small.

## 3. Superluminal laser

For a superluminal laser, the gain profile is a superposition of a broad gain and a narrow depletion. For concreteness of discussion, we consider the approach where the gain is produced in one isotope ($^{85}$Rb) via the Raman gain process and the depletion is produced in the other isotope ($^{87}$Rb) via the Raman depletion process, as shown in Figure 3(a) [20]. For the discussion in this paragraph, the only parts of Figure 3(a) that are relevant are the beams at frequencies $\omega_{P10}$, $\omega_{P20}$ and $\omega_{L0}$; the remaining parts of this figure are discussed later on. In $^{85}$Rb, an optical pump (not shown in the diagram) couples $|1\rangle$ to a state that is far away from state $|3\rangle$, so that its interaction with the $|1\rangle \leftrightarrow |3\rangle$ and the $|2\rangle \leftrightarrow |3\rangle$ transitions can be ignored. As such, the net effect of this pump is to produce an incoherent transfer of atoms from $|1\rangle$ to $|2\rangle$. When the Raman pump 1 (RP1) at frequency $\omega_{P10}$ is applied to the $|2\rangle \leftrightarrow |3\rangle$ transition, a Raman gain is produced for a probe which couples the $|1\rangle \leftrightarrow |3\rangle$ transition, with peak gain occurring at probe frequency $\omega_{L0}$. In $^{87}$Rb, a different optical pump is applied in opposite configuration and transfers atoms from $|2'\rangle$ to $|1'\rangle$ incoherently. Thus, the same probe field experiences Raman depletion around the two-photon resonance frequency in the presence of the Raman pump 2 (RP2) at frequency $\omega_{P20}$ applied to the $|2'\rangle \leftrightarrow |3'\rangle$ transition, with maximum depletion occurring at probe frequency $\omega_{L0}$. Thus, the center frequencies of the gain profile and the depletion profile occur at the same probe frequency. In order for this to happen, the necessary condition is that the frequency difference between the two Raman pumps must match the difference between the ground-state hyperfine splitting in $^{87}$Rb and the same in $^{85}$Rb. Explicitly, this means that $\omega_{P20} - \omega_{P10} = \omega_{21-87} - \omega_{21-85}$, where $\omega_{21-87}$ ($\omega_{21-85}$) is the energy difference between the two hyperfine ground states in $^{87}$Rb ($^{85}$Rb). The resulting gain profile is illustrated as the solid trace in Figure 3(b). In practice, RP2 is offset phase locked to RP1 or generated from RP1 utilizing an EOM (Electro-Optic Modulator). We assume that the cavity length is tuned to ensure that the frequency of one of the longitudinal modes is at frequency $\omega_{L0}$. Therefore, superluminal lasing will occur at this frequency, assuming that the gain at the bottom of the dip exceeds the loss, and that the dispersion produces a group index less than unity.

Consider now a situation when the frequency of RP1 changes to $\omega'_{P1}$, as illustrated in the left panel of Figure 3(a). The shift in the frequency of RP1 is denoted as $\Delta_P$. The frequency of RP2 is, by construction, shifted by the same amount, namely $\Delta_P$, as illustrated in the right panel of Figure 3(a). Therefore, the gain spectrum retains the same shape but is shifted in frequency by $\Delta_P$, which is illustrated as the dotted trace in Figure 3(b). The center frequencies of the gain profile and the depletion profile would now be $\omega'_L$, as illustrated in both panels of Figure 3(a). Explicitly, we have the condition that $\Delta_P = \omega'_{P1} - \omega_{P10} = \omega'_{P2} - \omega_{P20} = \omega'_L - \omega_{L0}$. As a result of the change in the gain profile, the frequency of the superluminal laser will move to a new value, denoted as $\omega_L$, as indicated by the green arrows in both panels of Figure 3(a). We define the shift in the frequency as $\Delta_L \equiv \omega_L - \omega_{L0}$. In what follows, we determine the value of this shift.

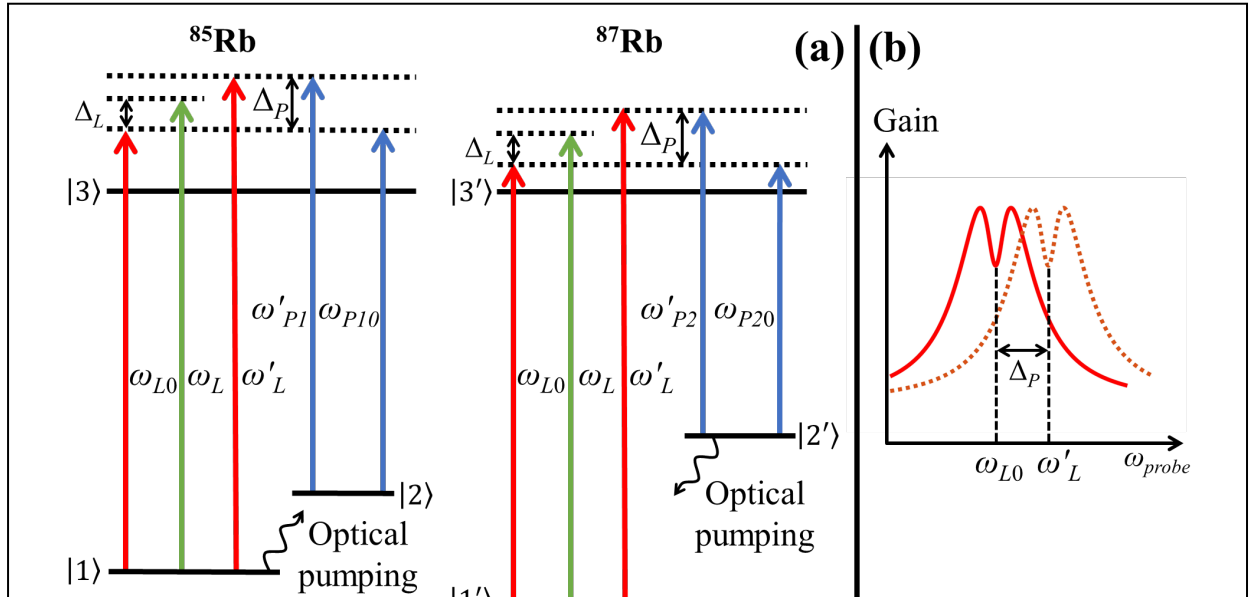

Figure 3. Schematic of (a) the optical fields and the relevant energy levels in the superluminal laser, and (b) the gain spectra as functions of the superluminal laser frequency before the Raman pump frequency shifts (solid line) and after the Raman pump frequency shifts by $\Delta_P$ (dotted line), respectively. See text for details.

As stated previously, the index after lasing for a subluminal laser is linear under the condition where $\Delta_P$ is smaller than the width of the unsaturated gain spectrum, which, in most cases, is satisfied within the lasing range shown in Figure 2(c). However, for the superluminal laser, the index is nonlinear due to the fact that the gain spectrum is a superposition of a broad gain and a narrow dip. The unsaturated gain and index can be written as [15]:

$$G_u = \frac{G_1\Gamma_1^2/2}{2\theta_1^2\Omega_1^2+\Gamma_1^2+4(\omega-\omega_{L0})^2} - \frac{G_2\Gamma_2^2/2}{2\theta_2^2\Omega_2^2+\Gamma_2^2+4(\omega-\omega_{L0})^2}, \quad (33)$$

$$n_u = 1 + \frac{G_1\Gamma_1(\omega-\omega_{L0})}{2\theta_1^2\Omega_1^2+\Gamma_1^2+4(\omega-\omega_{L0})^2} - \frac{G_2\Gamma_2(\omega-\omega_{L0})}{2\theta_2^2\Omega_2^2+\Gamma_2^2+4(\omega-\omega_{L0})^2}, \quad (34)$$

where the subscript $u$ indicates the unsaturated quantities, $G_{1(2)}$ and $\Gamma_{1(2)}$ are the maximum gain and the linewidth of the broad gain (narrow dip), $\Omega_{1(2)}$ is the Rabi frequency of the superluminal laser field for $^{85(87)}$Rb atoms, and $\theta_{1(2)} \equiv \Omega_{P1(P2)}/2|\delta_{P1(P2)}|$ with the Rabi frequency of RP1 (RP2) being denoted as $\Omega_{P1(P2)}$. In steady state, the gain experienced by the laser equals the loss of the cavity. Therefore, to evaluate the saturated gain and dispersion, Eq. (33) is set equal to the loss of the cavity, namely $1/2Q$. The Rabi frequencies as functions of the lasing frequency can be then determined. As established previously, we first express the Rabi frequency as:

$$\Omega_{1(2)}^2 = \frac{\varepsilon_0 G_{1(2)}\Gamma_{1(2)}}{2\hbar N_{1(2)}\theta_{1(2)}^2}E^2, \quad (35)$$

$$G_{1(2)} = \frac{2N_{1(2)}\theta_{1(2)}^2\mu_{1(2)}^2}{\hbar\varepsilon_0\Gamma_{1(2)}}, \quad (36)$$

where $N_{1(2)}$ is the number density of $^{85(87)}$Rb atoms, $\mu_{1(2)}$ is the dipole moment of a single item for the $|1\rangle \leftrightarrow |3\rangle$ ($|1'\rangle \leftrightarrow |3'\rangle$) transition, and $E$ is the amplitude of the superluminal laser electric field. The saturated electric field amplitude can be determined by solving the equation:

$$\frac{\zeta_1}{E^2+\eta_1} - \frac{\zeta_2}{E^2+\eta_2} = \frac{1}{2Q}, \quad (37)$$

$$\eta_{1(2)} \equiv \left[\Gamma_{1(2)}^2 + 4(\omega-\omega_{L0})^2\right]\frac{\hbar N_{1(2)}}{\varepsilon_0 G_{1(2)}\Gamma_{1(2)}}, \quad (38)$$

$$\zeta_{1(2)} \equiv \frac{\hbar N_{1(2)}\Gamma_{1(2)}}{2\varepsilon_0}. \quad (39)$$

It needs to be noted that $\eta_1$ and $\eta_2$ are functions of the frequency of the superluminal laser, while $\zeta_1$ and $\zeta_2$ are constants. Naturally, we keep the positive solution for $E^2$ to Eq. (37). As such, the square of the amplitude of the superluminal laser can be expressed as:

$$E^2 = -\frac{1}{2}\left[\eta_1+\eta_2-2Q(\zeta_1-\zeta_2)\right]+\frac{1}{2}\sqrt{\left[\eta_1-\eta_2-2Q(\zeta_1+\zeta_2)\right]^2-16Q^2\zeta_1\zeta_2}. \quad (40)$$

By substitute this solution into Eq. (34), the saturated refractive index can be determined. To be specific:

$$n_s = 1 + \frac{2\zeta_1(\omega - \omega_{L0})/\Gamma_1}{E^2 + \eta_1} - \frac{2\zeta_2(\omega - \omega_{L0})/\Gamma_2}{E^2 + \eta_2}, \quad (41)$$

where the subscription $s$ indicates the saturated quantity. It needs to be noted that in this expression, $E^2$ is a function of $(\omega - \omega_{L0})$ as well as $\eta_{1(2)}$. As a result, the saturated index is not simply a linear function, unlike the case of the subluminal laser. As such, the group index cannot not be conveniently written down in an analytical form.

Next, we consider the effect of changing the frequencies of the Raman pumps. In principle, the frequencies of the two Raman pumps could vary independently. However, as mentioned earlier, the frequencies of the two Raman pumps are off-set phase locked to one another, or the second Raman pump is generated from the first one by using an EOM. As such, any change in the frequency of one Raman pump would be the same as that of for the other Raman pump. We denote by $\Delta_P$ the change in the frequencies of both Raman pumps. The unsaturated gain and index can then be expressed as:

$$G'_u = \frac{G_1 \Gamma_1^2 / 2}{2\theta_1^2 \Omega_1'^2 + \Gamma_1^2 + 4(\omega - \omega'_L)^2} - \frac{G_2 \Gamma_2^2 / 2}{2\theta_2^2 \Omega_2'^2 + \Gamma_2^2 + 4(\omega - \omega'_L)^2}, \quad (42)$$

$$n'_u = 1 + \frac{G_1 \Gamma_1 (\omega - \omega'_L)}{2\theta_1^2 \Omega_1'^2 + \Gamma_1^2 + 4(\omega - \omega'_L)^2} - \frac{G_2 \Gamma_2 (\omega - \omega'_L)}{2\theta_2^2 \Omega_2'^2 + \Gamma_2^2 + 4(\omega - \omega'_L)^2}. \quad (43)$$

To derive the expression for the saturated index, we set Eq. (42) to $1/2Q$ and find the solution for $E'^2$ as:

$$E'^2 = -\frac{1}{2}\left[\eta'_1 + \eta'_2 - 2Q(\zeta_1 - \zeta_2)\right] + \frac{1}{2}\sqrt{\left[\eta'_1 - \eta'_2 - 2Q(\zeta_1 + \zeta_2)\right]^2 - 16Q^2 \zeta_1 \zeta_2}. \quad (44)$$

For simplicity in notations, we define:

$$\eta'_{1(2)} \equiv \left[\Gamma_{1(2)}^2 + 4(\omega - \omega'_L)^2\right] \frac{\hbar N_{1(2)}}{\varepsilon_0 G_{1(2)} \Gamma_{1(2)}}. \quad (45)$$

The saturated index, with a frequency shift $\Delta_P$ in the Raman pumps, can be written as:

$$n'_s = 1 + \frac{2\zeta_1(\omega - \omega'_L)/\Gamma_1}{E'^2 + \eta'_1} - \frac{2\zeta_2(\omega - \omega'_L)/\Gamma_2}{E'^2 + \eta'_2}. \quad (46)$$

The new lasing frequency needs to satisfy the resonance condition:

$$\omega_L \cdot n'_s \big|_{\omega = \omega_L} = \omega_{L0}. \quad (47)$$

As can be seen from Eqs. (44) through (46), $n'_s$ is a nonlinear function of $(\omega - \omega'_L)$. As such, it is not possible, for the general case, to solve Eq. (47) analytically, and a numerical approach has to be used to solve Eq. (47) and then evaluate the ratio $(\Delta_L / \Delta_P)$, as shown later.

A qualitative understanding of the expected behavior of the ratio $(\Delta_L / \Delta_P)$ can be obtained by considering the situation where the frequency shift of the superluminal laser is much smaller than the linewidth of the narrow depletion profile, i.e. $\Delta_L / \Gamma_2 \ll 1$. In this limit, the terms $(\omega_L - \omega'_L)/\Gamma_{1(2)}$ in Eq. (47) can be ignored. The saturated index can be then simplified as:

$$n'_s \approx 1 + \frac{2\zeta_1(\omega - \omega'_L)/\Gamma_1}{E'^2 + 2\zeta_1/G_1} - \frac{2\zeta_2(\omega - \omega'_L)/\Gamma_2}{E'^2 + 2\zeta_1/G_1} \qquad (48)$$
$$\equiv 1 + (\alpha' - \beta')(\omega - \omega'_L),$$

$$\eta'_{1(2)} \approx \frac{\hbar N_{1(2)} \Gamma_{1(2)}}{\varepsilon_0 G_{1(2)}} = \frac{2\zeta_{1(2)}}{G_{1(2)}}, \qquad (49)$$

$$E'^2 \approx -\left[\frac{\zeta_1}{G_1} + \frac{\zeta_2}{G_2} - Q(\zeta_1 - \zeta_2)\right] + \sqrt{\left[\frac{\zeta_1}{G_1} - \frac{\zeta_2}{G_2} - Q(\zeta_1 + \zeta_2)\right]^2 - 4Q^2 \zeta_1 \zeta_2}, \qquad (50)$$

with the following definitions for simplicity in notations:

$$\alpha' \equiv \frac{2\zeta'_1/\Gamma_1}{E'^2 + 2\zeta'_1/G_1}, \qquad (51)$$

$$\beta' \equiv \frac{2\zeta'_2/\Gamma_2}{E'^2 + 2\zeta'_2/G_2}. \qquad (52)$$

As can be seen, $\alpha'$, $\beta'$, and $E'^2$ are independent of $(\omega - \omega'_L)$. The saturated index is then reduced to a superposition of two linear functions. If we define:

$$\tilde{\alpha} \equiv \alpha' - \beta', \qquad (53)$$

then Eq. (48) has the same form as Eq. (24). It is thus easy to show that the behavior of the frequency shift for the superluminal laser as a function of the change in the pump frequency, in the limit $(\omega_L - \omega_{L0})/\Gamma_2 \ll 1$, is identical to that of the subluminal laser. To be specific, the ratio between the frequency shifts of the superluminal laser and the Raman pumps can be expressed as:

$$\frac{\Delta_L}{\Delta_P} = \frac{n_g - 1}{n_g}. \qquad (54)$$

$$n_g(\omega_{L0}) \equiv \left.\frac{\partial(n'_s\omega)}{\partial\omega}\right|_{\omega_{L0}} = 1 + \tilde{\alpha}\omega_{L0}. \tag{55}$$

If we consider the situation where $0 < n_g \ll 1$, we can write:

$$\frac{\Delta_L}{\Delta_P} \approx -\frac{1}{n_g}. \tag{56}$$

As can be seen, as long as $\Delta_L/\Gamma_2 \ll 1$ and the group index is close to zero, the frequency shift in the superluminal laser is magnified by the inverse of the group index comparing to the frequency shift in the Raman pumps and in the opposite direction.

Given that, in this limit, the form of $\Delta_L/\Delta_P$ as a function of the saturated group index is the same as that of the subluminal laser (i.e. $(n_g - 1)/n_g$), it is possible to consider the frequency shifts for both types of lasers in a unified manner, as illustrated in Figure 4. Here, we have plotted $\Delta_L$ as a function of inverse of the saturated group index, for a fixed value of $\Delta_P$, namely 1 Hz. The case of the subluminal laser is illustrated on the left panel, while that of the superluminal laser is illustrated on the right panel. As can be seen, the sign of the shift is opposite for these two cases. Furthermore, for the subluminal case, the maximum shift is never bigger than the shift in the pump frequency. On the other hand, for a superluminal laser, the amplitude of the maximum shift can be much larger than the shift in the pump frequency when the inverse of the saturated group index is very large. However, it needs to be kept in mind that for a superluminal laser, this behavior is valid only for a very small change in the pump frequency, and the behavior differs significantly when the pump frequency shift is much larger. The range of pump frequency shift for which Eq. (54) remain valid for a superluminal laser depends on various parameters, such as the linewidths of the narrow depletion profile and the broad gain profile, as discussed next.

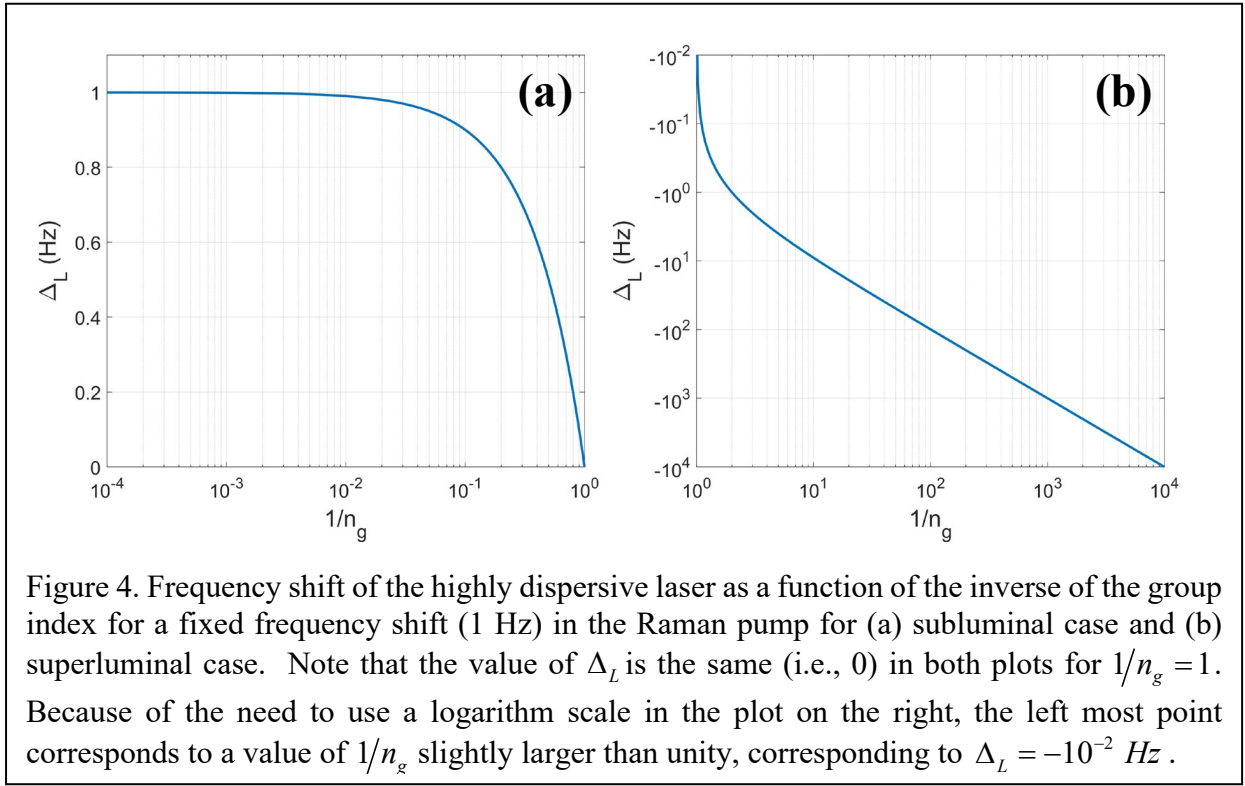

Figure 4. Frequency shift of the highly dispersive laser as a function of the inverse of the group index for a fixed frequency shift (1 Hz) in the Raman pump for (a) subluminal case and (b) superluminal case. Note that the value of $\Delta_L$ is the same (i.e., 0) in both plots for $1/n_g = 1$. Because of the need to use a logarithm scale in the plot on the right, the left most point corresponds to a value of $1/n_g$ slightly larger than unity, corresponding to $\Delta_L = -10^{-2}$ Hz.

To investigate the effect of the nonlinearity of the refractive index, we choose fixed values of $G_1$, $\Gamma_1$, and $\Gamma_2$, and vary the amplitude, $G_2$, of the narrow depletion profile to achieve different group indices. Equation (47) is then solved numerically to find the lasing frequencies when the frequencies of the Raman pumps are shifted by different values. Specifically, we make use of the solution of $E'^2$ shown in Eq. (44) and substitute it in Eq. (46) to find the saturated index as a function of $(\omega - \omega_L')$. We then solve Eq. (47) by numerically evaluating $\varepsilon \equiv \left(\omega_L \cdot n_s'\big|_{\omega=\omega_L} - \omega_{L0}\right)$ with a varying value of $\omega_L$. The solution to Eq. (47) is deemed to be accurate enough when a specific value of $\omega_L$ yields a value of $\varepsilon$ smaller than the chosen convergence tolerance. This process is repeated for different values of $n_g$ and $\Delta_P$. The results are illustrated in Figure 5, which shows the value of $\Delta_L/\Delta_P$ as a function of the inverse of the saturated group index, for different value of $\Delta_P$. As can be seen, the frequency of the superluminal laser is shifted in the direction opposite to that of the frequencies of the Raman pumps (i.e. the value of $\Delta_L/\Delta_P$ is negative). The parameters for generating Figure 5 are as follows. The maximum gain and the linewidth of the broad gain profile are respectively $1.2\times10^5$ and $2\pi\times30$ MHz. The linewidth of the narrow depletion profile is $2\pi\times10$ MHz. $N_1$ and $N_2$ are $9\times10^6$ m$^{-3}$ and $1\times10^{11}$ m$^{-3}$, respectively. $Q$ is set to be $10^6$. When the superluminal laser frequency shift is small compared to the linewidth of the narrow depletion profile, the frequency shift in the superluminal laser agrees very closely with the result shown in Eq. (54). Specifically, when the inverse of the saturated group index is very large, the value of $\Delta_L/\Delta_P$ is approximately $-1/n_g$. On the other hand, when the inverse of the saturated group index is comparable to unity, the value of $\Delta_L/\Delta_P$ is approximately $(n_g-1)/n_g$. For larger values of $\Delta_P$, the ratio $|\Delta_L/\Delta_P|$ approaches a smaller value asymptotically with increasing values of the inverse of the saturated group index. This is due to the fact that when the value of $\Delta_L$ is outside the linear region in the derivative of the Lorentzian profile, the superluminal laser experiences a saturated group index that is much larger than that within the linear region, yielding a smaller magnification factor. As such, in a superluminal laser, a large ratio of $|\Delta_L/\Delta_P|$ occurs under two conditions. One is when the frequency shifts in the Raman pumps are significantly smaller than the linewidth of the narrow depletion profile and the other is when the saturated group index at the superluminal laser frequency is close to zero.

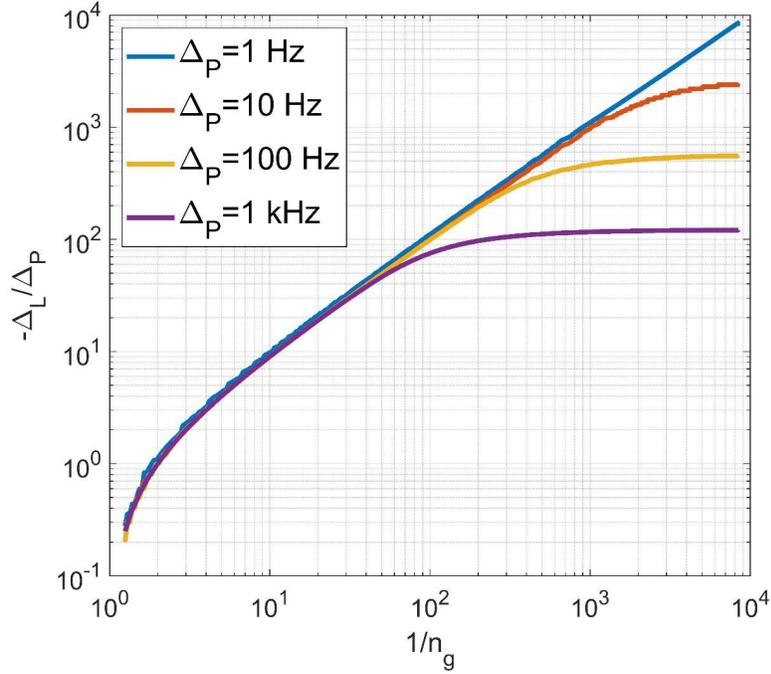

Figure 5. The ratio of $-(\Delta_L/\Delta_P)$ in a superluminal laser as functions of the inverse of the group index for different values of the frequency shift in the Raman pumps.

To validate all the approximations employed when deriving Eq. (48), we solve the laser frequency using the explicit three-level systems in each isotope that produce the superluminal dispersion spectrum. Specifically, we solve the density matrix equations of motion for the three-level system for each isotope and the single mode laser equation simultaneously in an iterative manner [13,18]. We first find the lasing frequency for the original Raman pump frequency and a certain group index. Then we change the Raman pump frequency by $\Delta_P$, execute the algorithm, and find the new lasing frequency and the corresponding change, $\Delta_L$. This process is repeated for different values of the group index. The results are compared with the approximated Lorentzian model, presented in Eq. (33) and Eq. (34), for the superluminal laser case for the same parameters and $\Delta_P = 1\,\text{Hz}$. Figure 6 shows the ratio $\Delta_L/\Delta_P$ as functions of group index for the system involving three-level systems in each isotope, illustrated as solid trace, and the approximated expression, illustrated as the dashed trace. The small variations in the dual isotopes three-level result are caused by numerical errors, which can be suppressed by reducing the convergence tolerance and increasing the frequency resolution. However, this process would increase the computation time drastically. The degree of agreement shown in Figure 6 is arguably adequate enough for validating the approximations employed earlier. It should be noted that a similar investigation can also be carried out to determine the frequency shift for a subluminal laser using a density matrix analysis of an optically pumped three-level system in a single isotope. However, we have chosen not to carry out such a study since the analytic results found for that case are far simpler than those for the superluminal laser, and valid over a large range of parameters.

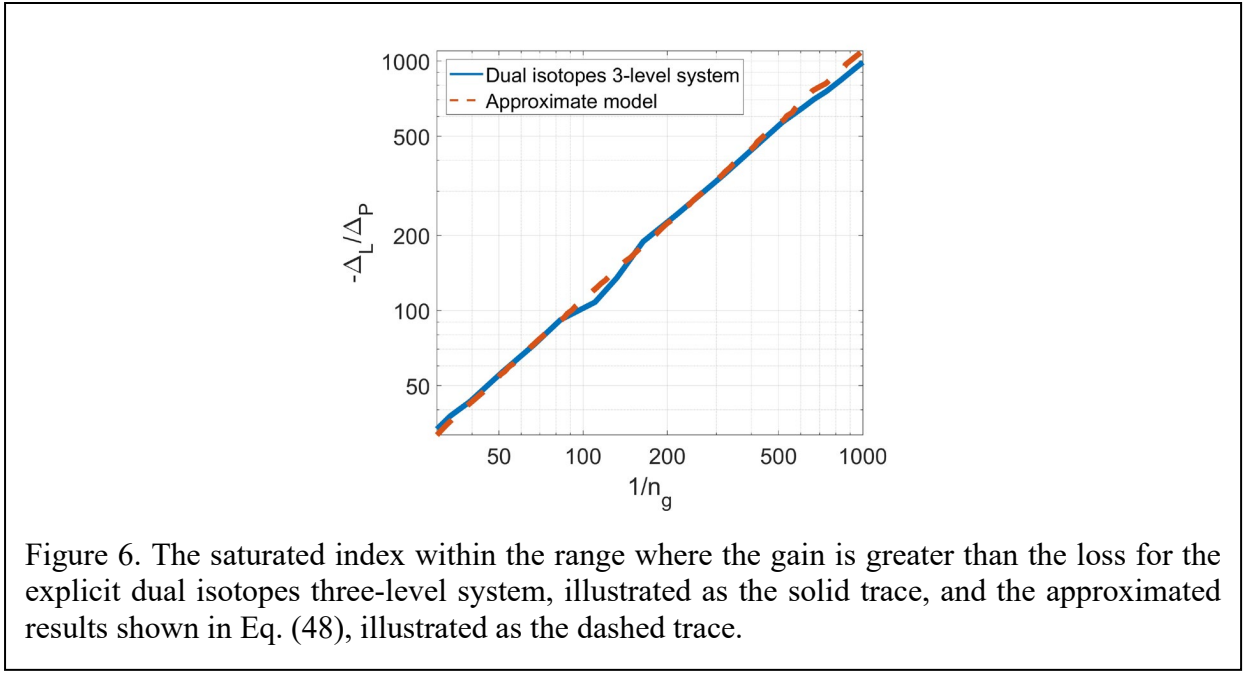

Figure 6. The saturated index within the range where the gain is greater than the loss for the explicit dual isotopes three-level system, illustrated as the solid trace, and the approximated results shown in Eq. (48), illustrated as the dashed trace.

## 4. Effects of Light-shifts

In the analyses presented above, we have calculated the frequency shift in highly dispersive lasers (HDLs) caused by a shift in the Raman pump frequency for the case of a subluminal laser, and a matching shift in the frequencies of both Raman pumps for the case of a superluminal laser. Here, we discuss the effects of light-shifts of the energy levels of the ground hyperfine states and variations thereof on the HDL frequency. The light-shifts are produced by the Raman pumps, the HDL itself, and the optical pumps. In what follows, for the sake of simplicity, we will refer to a single Raman pump, since the subluminal laser uses only one Raman pump, and the effects of each of the two Raman pumps for the superluminal laser are similar. Similarly, we will refer to a single optical pump, since the subluminal laser uses only one optical pump, and the effects of each of the two optical pumps for the superluminal laser are similar.

      Consider first the light-shift produced by the Raman pump. The following points need to be noted in this context. First, the light-shift corresponding to the initial frequency of the Raman pump is implicitly incorporated into the effective value of the two-photon detuning prior to reducing the three-level system to the effective two-level system. Second, a change in the frequency of the Raman pump, $\Delta_P$, also changes the value of the light-shift, which should be taken into account in determining the shift in the unsaturated gain profile. However, compared to

the amount of the center frequency shift in the gain profile directly caused by $\Delta_P$, this effect is negligible, since the Raman pump is already highly detuned with respect to the optical transition.

Consider first the light-shift produced by the HDL itself. The following points need to be noted in this context. First, the light-shift corresponding to the initial frequency of the HDL is already taken into account implicitly, in a self-consistent manner, when the cavity is tuned to the length that makes the HDL operate at the center of the gain peak (dip) for the subluminal (superluminal) case. Second, a change in the frequency of the HDL would also change the value of the light-shift, which should be taken into account in determining the shift in the saturated gain profile. However, compared to the amount of the HDL frequency shift, any additional change in its frequency due to this effect is negligible, since it is already highly detuned with respect to the optical transition.

Additionally, it should be noted that the density matrix based numerical analysis naturally takes into account the effects of light-shifts induced by the Raman pump and the HDL. As such, predictions made using this approach are expected to be fundamentally more accurate. However, as can be seen from Figure 6, the agreement between this model and the analytical model is still quite good, validating our observation that the effect of variations in the light-shifts due to the Raman pump and the HDL is negligible.

Consider next the effect of light-shift induced by the optical pumping laser. For specificity, we consider the case where the HDL operates along the D1 transition, so that the optical pumping laser (OPL) couples one of the ground hyperfine states to the hyperfine states in the $5P_{3/2}$ manifold. The following points should be noted in this context. First, the OPL frequency can be tuned to an optimal value so that light-shift averaged over each of the upper-level hyperfine states and the Doppler shifts has a null value. The fact that this is possible is evident from the observation that the net light shift from the OPL changes sign when its frequency is scanned from a value below resonance with respect to the $5P_{3/2}$ manifold to one that is above resonance. The OPL frequency can be stabilized to this optimal value by using an AOM to shift the frequency of a part of the OPL and locking this shifted frequency to one of the hyperfine transitions. Second, if we make the reasonable assumption that the Raman pump and the OPL have similar degrees of frequency stability, then the OPL induced light shift would be negligibly small when operating around this optimal frequency. Finally, it should be noted that in the three-level model(s) presented above, we have used an effective decay rate to account for the effect of the optical pump. However, it is easy to update the model to include another energy level used for the optical pumping transition [22]. If the density matrix based model is used with such a four-level model, then all the effects of light shifts from the optical pumping process would be fully accounted for automatically.

In the preceding discussions we have not included the effect of changes in the light-shifts caused by power variations of the Raman pumps and the OPLs. Of course, such changes would be formally equivalent to changes in the frequencies of these lasers. The findings presented here can thus be used to estimate the degree of intensity stability of the pumps that may be needed for a given application. Such a study would be carried out in the near future.

## 5. Implications for Sensors Employing Highly Dispersive Lasers

While many applications can be envisioned for the HDLs, arguably the most obvious one is the use of a superluminal laser for ultra-sensitive rotation sensing. A superluminal ring laser gyroscope (SRLG) would have two counter-propagating superluminal lasers operating in the same cavity. The findings of the analyses presented above show that care must be taken in designing such an SRLG to suppress the effects other than rotation that can lead to differential variations in the frequencies of the two superluminal lasers. Specifically, the same Raman pumps and optical pumps must be used for both directions, and the intensities of these pumps in both directions should also be as closely matched as possible.

Another possible application of a superluminal laser is to measure acceleration or vibration using a unidirectional superluminal ring laser (SRL) with one of its mirrors mounted on a metallic diaphragm or a spring. For such a device, the findings reported here show that the ultimate sensitivity of the device would be constrained by the frequency stabilities of the Raman pump lasers. This constraint can be circumvented by making use of two co-propagating SRLs in different but matching cavities, with one placed above the other, and using an acceleration sensitive mirror in only one of the cavities. If both of these SRLs make use of the same Raman pumps and optical pumps, with matching power levels, then the frequency difference between the two SRLs would yield a signal proportional to the acceleration. This configuration would have the added benefit of being insensitive to rotation.

Broadly speaking, for any application of the HDL, one must take into account the findings reported in this paper. In principle, one can make use of Raman pump lasers and optical pump lasers that are sufficiently stable for the application at hand. However, for some applications, such detection of ultra-light scalar field dark matter [11,12], one must also take into account the fact that the frequencies of the pump lasers would also vary due to the dark matter. Similarly, for frequency shift measurement based gravitational wave detection, it would be necessary to account for any frequency shift of the pump lasers induced by the gravitational waves.

## 6. Conclusions

In this paper, we have studied how the frequency of a highly dispersive laser (HDL) based on Raman gain (and Raman depletion) changes due to changes in frequency(-ies) of the Raman pump laser(s). Specifically, we have investigated how the degree of change in the HDL frequency depends on the group index experienced by the intracavity laser field. If the HDL is highly subluminal, the output frequency shift converges to the value of the frequency shift in the Raman pump laser. On the contrary, if the HDL is highly superluminal, the ratio of the HDL frequency shift and the Raman pump frequency shift is negative, and its amplitude equals the inverse of the group index, when the HDL frequency shift is small compared to the linewidth of the depletion profile. When this constraint is not met, the ratio saturates at some value of the inverse of the group index, and the saturation point depends on a combination of various parameters, such as the widths of the gain profile and the depletion profile. We have also shown that related effects due to light-shifts and changes thereof produced by the Raman pump(s) and the optical pump(s), as well as the HDL itself are expected to be negligible. The finding reported here must be taken into account in designing sensors based on HDLs.

**Acknowledgement:**

This work has been supported by AFOSR (FA9550-18-01-0401 and FA9550-21-C-0003), NASA (80NSSC22CA052), and Defense Security Cooperation Agency (PO4441028735).

## Appendix A: Effective Two-Level Model for a Raman Laser

For the 3-level system illustrated in the left panel of Figure 1, the complex Hamiltonian with the decay rate, after applying rotating wave transformation and rotating wave approximation, can be written as:

$$\tilde{H} = \hbar \begin{bmatrix} \delta_{diff} - i\Gamma_{eff}/2 & 0 & \Omega_L/2 \\ 0 & 0 & \Omega_P/2 \\ \Omega_L/2 & \Omega_P/2 & -\delta_P - i\Gamma_3/2 \end{bmatrix}. \tag{57}$$

where the quantities are defined in the caption of Figure 1. The amplitude equations can be then written as:

$$\dot{C}_1 = -i\left(\delta_{diff} - i\Gamma_{eff}/2\right)C_1 - i\left(\Omega_L/2\right)C_3, \tag{58}$$

$$\dot{C}_2 = -i\left(\Omega_P/2\right)C_3, \tag{59}$$

$$\dot{C}_3 = -i\left(\Omega_L/2\right)C_1 - i\left(\Omega_P/2\right)C_2 + i\left(\delta_P + i\Gamma_3/2\right)C_3, \tag{60}$$

where $C_i$, with $i = 1, 2, 3$, is the amplitude of the $i$-th energy level. While these equations account for the decay of states $|1\rangle$ and $|3\rangle$, they do not incorporate the effect of influx of atoms to different states resulting from this decay. However, under conditions where the population of $|3\rangle$ remains very small (either because of very high common mode detuning or very small Rabi frequencies), ignoring of influx of atoms to states $|1\rangle$ and $|2\rangle$ from state $|3\rangle$ remains a good approximation, as can be shown by comparing the solution found from the effective two level model with the exact solution of the full density matrix equations. On the other hand, the influx of atoms from state $|1\rangle$ to state $|2\rangle$ cannot be ignored. However, this will be taken into account when the amplitude equations representing the effective two level system is augmented by the effect of the influx from state $|1\rangle$ to state $|2\rangle$ to produce the corresponding density matrix equations of motion.

In the case relevant to this paper, the two fields are far detuned with respect to the $|1\rangle \leftrightarrow |3\rangle$ and the $|2\rangle \leftrightarrow |3\rangle$ transitions. As such, the excitation to state $|3\rangle$ is vanishingly weak, and we can assume that $\dot{C}_3$ approximately equals to zero. We can then express $C_3$ in terms of $C_1$ and $C_2$ as:

$$C_3 \approx \frac{\Omega_L}{2\delta_P + i\Gamma_3}C_1 + \frac{\Omega_P}{2\delta_P + i\Gamma_3}C_2. \tag{61}$$

For the Raman gain process, we have the assumption that $|\delta_P| \gg \Gamma_3$. Under this condition, Eq. (61) can be approximately reduced to:

$$C_3 \approx \frac{\Omega_L}{2\delta_P}C_1 + \frac{\Omega_P}{2\delta_P}C_2. \tag{62}$$

Substituting $C_3$ in Eq. (58) and Eq. (59) with Eq. (62), we have:

$$\dot{C}_1 = -i\left(\delta_{diff} + \frac{\Omega_L^2}{4\delta_P} - i\Gamma_{eff}/2\right)C_1 - i\frac{\Omega_L\Omega_P}{4\delta_P}C_2, \tag{63}$$

$$\dot{C}_2 = -i\frac{\Omega_L\Omega_P}{4\delta_P}C_1 - i\frac{\Omega_P^2}{4\delta_P}C_2. \tag{64}$$

These equations can be represented by the following effective Hamiltonian for the two level system involving only states $|1\rangle$ and $|2\rangle$:

$$\tilde{H}' = \hbar \begin{bmatrix} \left(\delta_{diff} + \frac{\Omega_L^2}{4\delta_P} - i\frac{\Gamma_{eff}}{2}\right) & \frac{\Omega_L\Omega_P}{4\delta_P} \\ \frac{\Omega_L\Omega_P}{4\delta_P} & \frac{\Omega_P^2}{4\delta_P} \end{bmatrix}. \tag{65}$$

Next, we apply another transformation to set the energy of $|1\rangle$ to be zero, and the resulting Hamiltonian becomes:

$$\tilde{\tilde{H}}' = \hbar \begin{bmatrix} -i\frac{\Gamma_{eff}}{2} & \frac{\Omega_L\Omega_P}{4\delta_P} \\ \frac{\Omega_L\Omega_P}{4\delta_P} & -\delta_{diff} - \frac{\Omega_L^2 - \Omega_P^2}{4\delta_P} \end{bmatrix}. \tag{66}$$

For compactness of notation, we define:

$$\Omega_{eff} \equiv \frac{\Omega_L\Omega_P}{4\delta_P}, \tag{67}$$

$$\delta \equiv \delta_{diff} + \frac{\Omega_L^2 - \Omega_P^2}{4\delta_P}. \tag{68}$$

The term $\left(\Omega_L^2 - \Omega_P^2\right)/4\delta_P$ in this equation is the difference in light shifts produced by the two fields. For the superluminal laser in steady state, this can be treated as essentially a constant. The Hamiltonian is then simplified as:

$$\tilde{\tilde{H}}' = \hbar \begin{bmatrix} -i\Gamma_{eff}/2 & \Omega_{eff}/2 \\ \Omega_{eff}/2 & -\delta \end{bmatrix}. \tag{69}$$

To incorporate the effect of the influx of atoms from state $|1\rangle$ to $|2\rangle$, we now form the corresponding density matrix equations:

$$\frac{\partial}{\partial t}\rho = -\frac{i}{\hbar}\left[\tilde{H}'\rho - \rho\tilde{H}'^{\dagger}\right] + \left(\frac{\partial \rho}{\partial t}\right)_{source}, \tag{70}$$

$$\rho = \begin{bmatrix} \rho_{11} & \rho_{12} \\ \rho_{21} & \rho_{22} \end{bmatrix}, \tag{71}$$

$$\left(\frac{\partial \rho}{\partial t}\right)_{source} = \begin{bmatrix} 0 & 0 \\ 0 & \Gamma_{eff}\rho_{11} \end{bmatrix}, \tag{72}$$

Since the population of state $|3\rangle$ is negligible, we can impose the closed-system constraint that $\rho_{11} + \rho_{22} = 1$. The steady state solution of Eq. (70) then yields:

$$\rho_{11} = \frac{\Omega_{eff}^2}{2\Omega_{eff}^2 + \Gamma_{eff}^2 + 4\delta^2} = 1 - \rho_{22}, \tag{73}$$

$$\rho_{21} = \frac{\Omega_{eff}(2\delta - i\Gamma_{eff})}{2\Omega_{eff}^2 + \Gamma_{eff}^2 + 4\delta^2} = \rho_{12}^*, \tag{74}$$

To estimate the coherence of the $|1\rangle \leftrightarrow |3\rangle$ transition, we multiply Eq. (62) by $C_1^*$, which yields:

$$\rho_{31} \approx \frac{\Omega_L}{2\delta_P}\rho_{11} + \frac{\Omega_P}{2\delta_P}\rho_{21}. \tag{75}$$

This is valid only when $|\rho_{21}| \approx \sqrt{\rho_{11}\rho_{22}}$, corresponding to a nearly pure state for the $|1\rangle \leftrightarrow |2\rangle$ coherence. One can use Eq. (73) and Eq. (74) to determine the parameter values for which this condition is satisfied, as shown below.

$$\frac{|\rho_{21}|}{\sqrt{\rho_{11}\rho_{22}}} = \frac{\sqrt{|\Omega_{eff}(2\delta - i\Gamma_{eff})|^2}}{\sqrt{\Omega_{eff}^2(\Omega_{eff}^2 + \Gamma_{eff}^2 + 4\delta^2)}}$$

$$= \sqrt{\frac{\Gamma_{eff}^2 + 4\delta^2}{\Omega_{eff}^2 + \Gamma_{eff}^2 + 4\delta^2}} \approx 1. \tag{76}$$

As can be seen, the condition $|\rho_{21}| \approx \sqrt{\rho_{11}\rho_{22}}$ is equivalent to $\Omega_{eff} \ll \Gamma_{eff}$.

By combining Eq. (75) with Eq. (73) and Eq. (74), the coherence of the $|1\rangle \leftrightarrow |3\rangle$ transition can be expressed as:

$$\rho_{31} \approx \frac{\Omega_{eff}}{2\delta_P \left(2\Omega_{eff}^2 + \Gamma_{eff}^2 + 4\delta^2\right)}\left[\Omega_L \Omega_{eff} + \Omega_P \left(2\delta - i\Gamma_{eff}\right)\right]. \quad (77)$$

Under the assumption $|\delta_P| \gg \Omega_P \gg \Omega_L$ for the unsaturated case, recalling Eq. (67), it is easy to show that $\Omega_L \Omega_{eff} \ll \Omega_P \Gamma_{eff}$. As such, the term $(\Omega_L / 2\delta_P)\rho_{11}$ in Eq. (75) can be ignored. As such, the expression for $\rho_{31}$ can be simplified as:

$$\rho_{31} \approx \frac{\Omega_P}{2\delta_P}\rho_{21} = \theta \rho_{21}, \quad (78)$$

where we have used the parameter $\theta$ defined in Eq. (34).

In Figure 7, we show comparison of the real and imaginary parts of $\rho_{31}$ found via explicit solution of the complete set of equations for the three-level system and the approximate result shown in Eq. (78). As can be seen, the agreement is excellent, thus justifying the approximations used in Eq. (78).

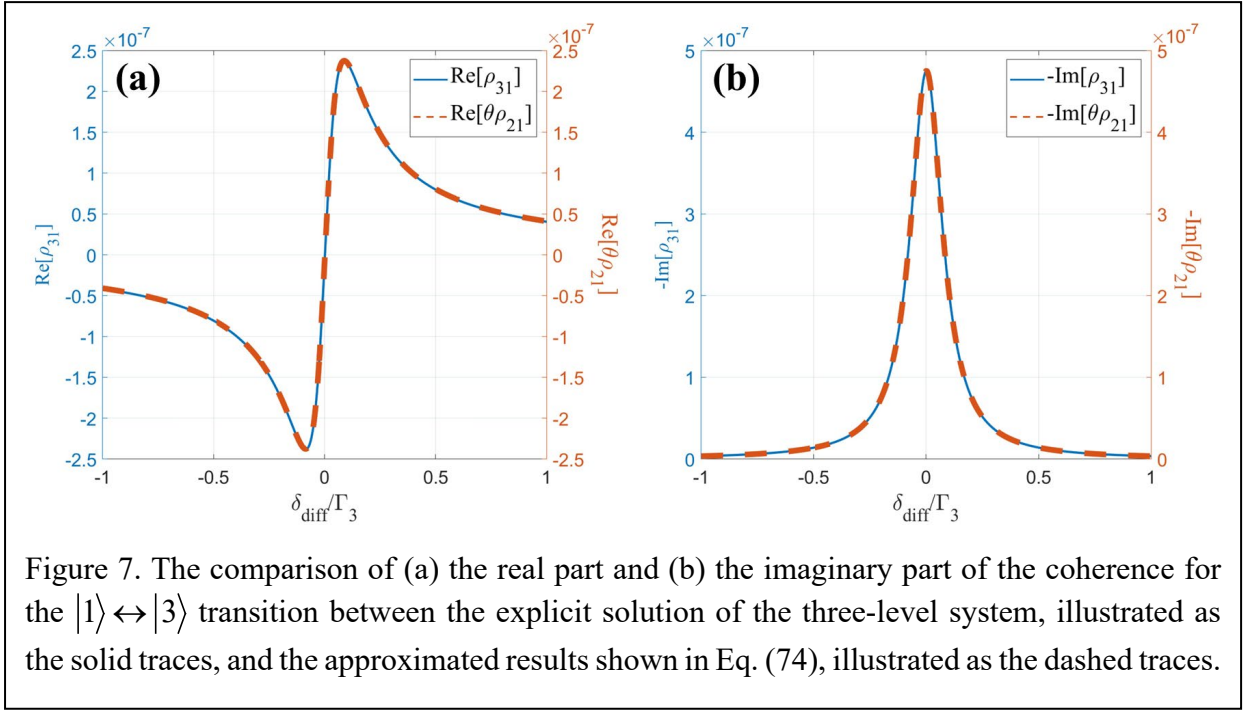

Figure 7. The comparison of (a) the real part and (b) the imaginary part of the coherence for the $|1\rangle \leftrightarrow |3\rangle$ transition between the explicit solution of the three-level system, illustrated as the solid traces, and the approximated results shown in Eq. (74), illustrated as the dashed traces.

For the saturated case, the Rabi frequency of the Raman laser is in general much larger than that in the unsaturated case. In order to validate the assumption $\delta_P \gg \Omega_P \gg \Omega_L$ under the saturated condition, it is necessary to evaluate $\Omega_L$ using Eq. (7) and Eq. (19):

$$\Omega_L^2 = \frac{\mu_0^2 E^2}{\hbar^2} = \frac{\mu_0^2}{\hbar^2}(2Q\zeta - \eta)$$
$$= \frac{\mu_0^2}{\hbar^2}\left[\frac{2\hbar N\Gamma Q}{\varepsilon_0} - (\Gamma_{eff}^2 + 4\delta^2)\frac{\hbar N}{\varepsilon_0 G_0 \Gamma_{eff}}\right]. \quad (79)$$

The maximum value of $\Omega_L$ appears when $\delta = 0$, so that we can rewrite Eq. (79) as:

$$\Omega_L^2\big|_{max} = \frac{\mu_0^2}{\hbar^2}\left(\frac{2\hbar N\Gamma_{eff}Q}{\varepsilon_0} - \frac{\hbar N\Gamma_{eff}}{\varepsilon_0 G_0}\right)$$
$$= \frac{\mu_0^2 N\Gamma_{eff}}{\hbar\varepsilon_0}(2Q - 1/G_0) \quad (80)$$

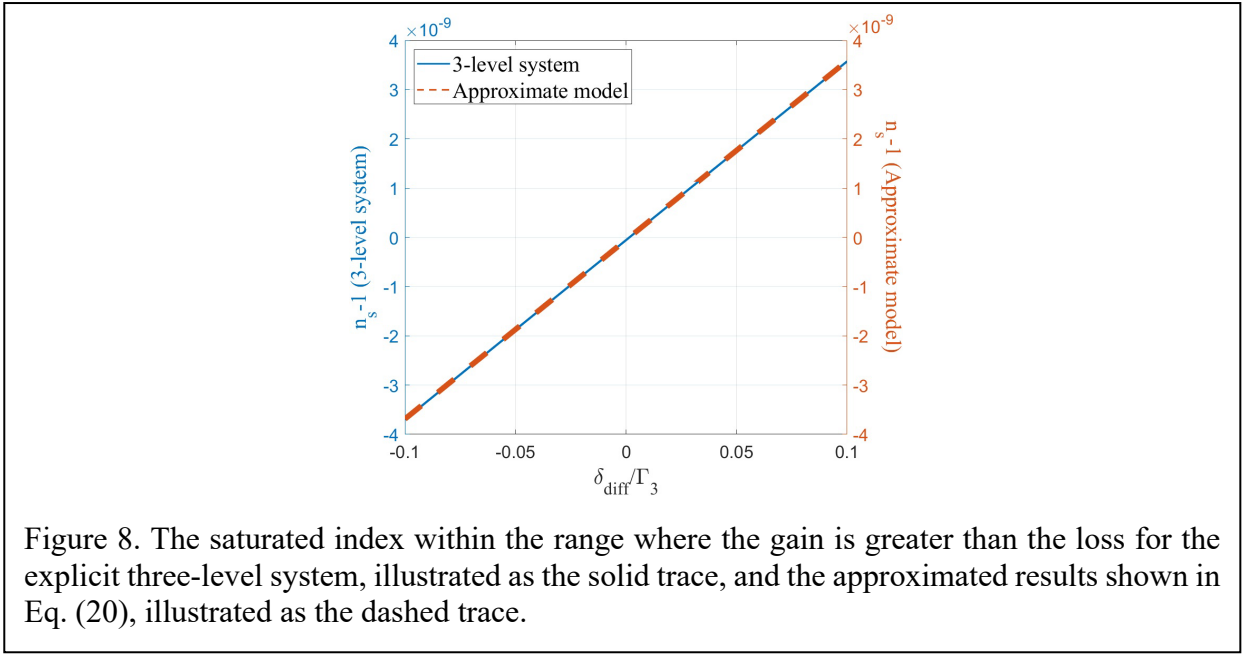

Figure 8. The saturated index within the range where the gain is greater than the loss for the explicit three-level system, illustrated as the solid trace, and the approximated results shown in Eq. (20), illustrated as the dashed trace.

As an example, consider the case where the peak gain, $G_0$, is twice the cavity loss rate, i.e. $G_0 = 1/Q$. Furthermore, we make use of the following values of the relevant parameters: $\mu_0 = 2.53 \times 10^{-29}\ C \cdot m$, corresponding to the dipole moment of the cycling transition in the D2 line of $^{85}$Rb, $N = 10^{16}\ m^{-3}$, $\Gamma_{eff} = 10^6\ rad/\sec$. Assume that the loss in the cavity is solely due to the finite transmission of the output coupler, the cavity quality factor is calculated using the following expression :

$$Q = \frac{2\pi L/\lambda}{1-R}, \quad (81)$$

where $L$ is the length of the cavity, which is chosen to be 0.1 $m$, $\lambda$ is the wavelength of the D2 line transition, which is 780 $nm$, and $R$ is the reflectivity of the output coupler of the cavity, which is set to be 95%. Using these parameter values in Eq. (80), we get :

$$\Omega_L^2 \Big|_{max} = \frac{\mu_0^2 N \Gamma}{\hbar \varepsilon_0} Q \approx \left(7.65 \times 10^7 \; rad/\sec\right)^2. \tag{82}$$

The detuning of the Raman pump, $|\delta_P|$, is usually more than 1 GHz for optimal Raman gain, which is equivalent to $6.28 \times 10^9 \; rad/\sec$. As can be seen, the assumption $|\delta_P| \gg \Omega_L$ holds for the saturated condition. Figure 8 illustrates the saturated index as a functions of $\delta_{diff}$ for the explicit solution of the three-level system (solid trace) and the approximate result using Eq. (20) (dashed trace), showing close agreement. Thus, the approximation in Eq. (78) is valid for both unsaturated and saturated cases.